\input{aipcheck}

\documentclass[
    ,final            
  ]
  {aipproc}

\layoutstyle{6x9}

\begin{document}

\newcommand{\Qt}{{\tilde Q}}
\newcommand{\X}{X}
\newcommand{\ba}{\begin{array}}
\newcommand{\ea}{\end{array}}
\newcommand{\dsp}{\displaystyle}

\title{Cold quark matter in astrophysics of compact stars}


\author{Armen Sedrakian}{
  address={Institute for Theoretical Physics,
J. W. Goethe-University, D-60438 Frankfurt am Main, Germany}
}

\begin{abstract}
The appearance of quark matter in the centers of 
compact stars has a number of astrophysical implications.
In this contribution I discuss the structure and stability of compact 
stars, gravitational radiation from non-axisymmetric 
deformations, and  nucleation and dynamics of vortices 
(flux tubes) in the color superconducting phases.
\end{abstract}

\maketitle


\section{Introduction}

The deconfinement phase transition in dense matter,  a manifestation 
of asymptotic freedom of quantum chromodynamics (QCD), may lead 
to stable quark matter cores in massive compact stars. The deconfined 
(``liberated'') quarks occupy continuum of scattering states, which in 
the low-temperature and high-density regime fill in the modes in a 
Fermi sphere. The behavior of quark matter  in this regime 
resembles that of less exotic low-temperature systems 
found in condensed matter 
({\it e.g.} electron gas or ultracold atomic vapor) or hadronic physics 
({\it e.g.} nuclear or neutron matter)~\cite{pairing_book}. 
In analogy to these system the 
attractive interaction between quarks, mediated by the gluon exchange,
which is responsible for the bound state spectrum of QCD, leads to quark 
superconductivity and superfluidity. This phenomenon is  commonly refered 
to as color superconductivity~\cite{Bailin:1983bm,Alford:2007xm}.
Interestingly, the inhomogeneous superconducting phases originally 
proposed in condensed matter and nucleonic system find their natural 
realization in compact stars when the $\beta$-equilibrium condition is 
imposed~\cite{Casalbuoni:2003wh}.

In the following, I will review and summarize 
some recent results concerning the astrophysical implications 
of the hypothesis of deconfinement and
color superconductivity in quark matter. 
This contribution is not intended to be a 
comprehensive review of the field and is 
centered mainly upon my recent research.

\section{Equilibrium and stability}
\label{sec:equilibrium}

While we are confident that neutron stars, which are observed 
as isolated pulsars or X-ray sources in binaries, contain nuclear 
matter in their interiors, the existence of hybrid stars with central 
cores made of quark matter is not obvious. Qualitatively, the uncertainty 
arises because deconfined quark matter has larger number of degrees
of freedom (color) than the ordinary nuclear matter and, consequently, 
its formation leads to a softening of the equation of states. Such softening 
may be fatal to the gravitational stability of the whole system; in fact, in 
many models of quark matter equation of state no stable configurations
are found -- the stars collapse into a black hole. A less severe 
uncertainty arises from the unknown physics of deconfinement and the density 
at which it occurs. The central densities of massive neutron stars are
at about five times the nuclear saturation density or more. The nucleons
will hardly remain intact at such high densities and the most plausible 
outcome of compression is a quark-gluon-fluid.

\begin{figure}[tb]
\includegraphics[height=.3\textheight,width=7.cm]{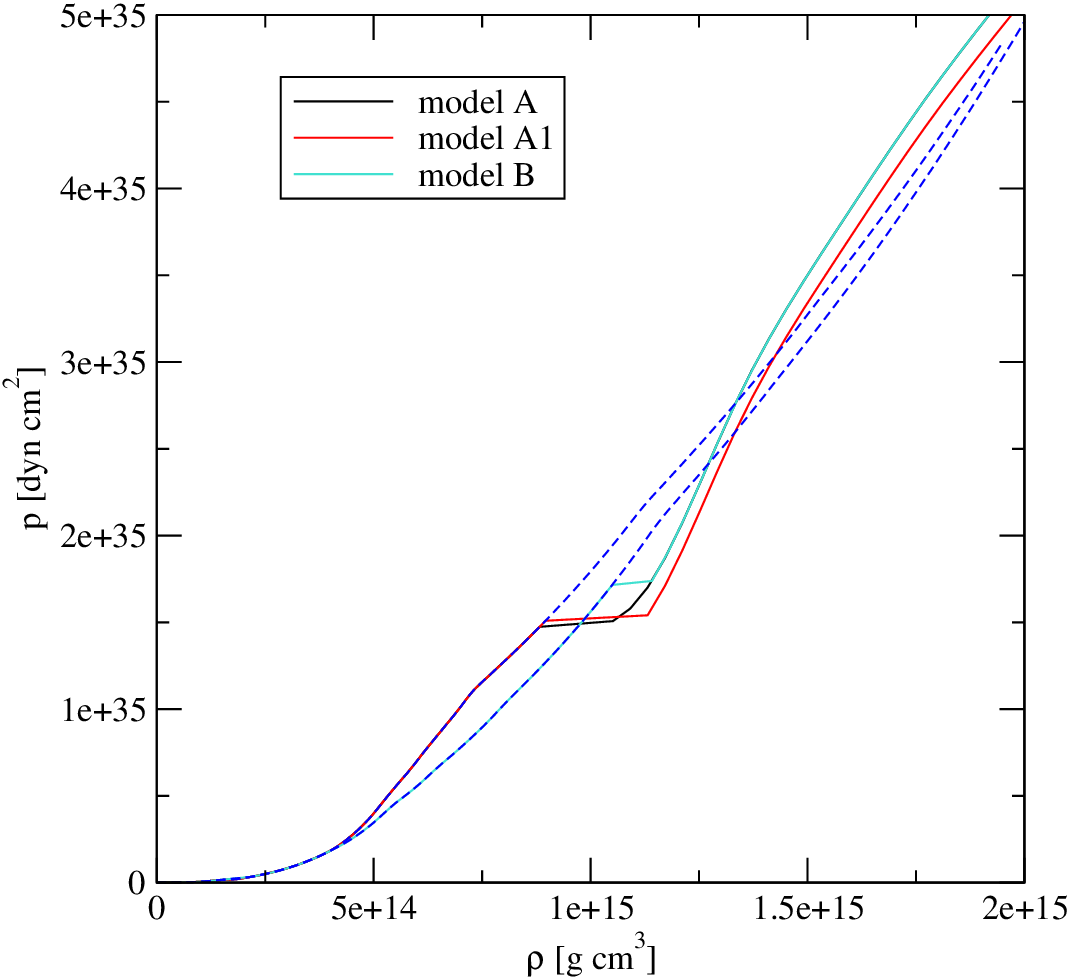}
\hskip 0.2cm
\includegraphics[height=.3\textheight,width=7.cm]{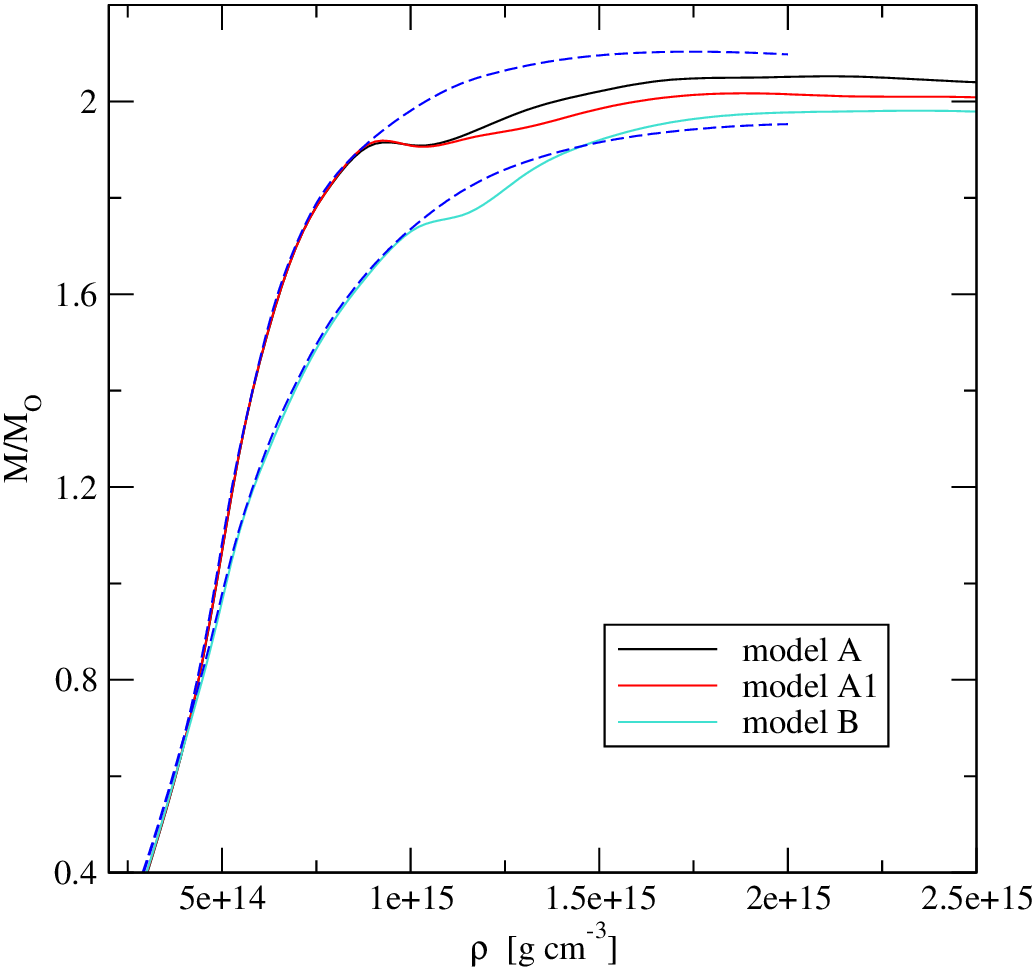}
\caption{
{\it Left panel}.
Number density versus pressure for three models.
For the models A and A1 the  nuclear (low density) equation of
state is the same; for the models A and B the  quark (high density)
equation of state is the same. {\it Right panel}.
Dependence of the total stellar mass and the mass of the quark core
in units of solar mass $M_{\odot}$
on the central density for non-rotating configurations. 
The lower set of curves represents the masses of the 
CCS quark cores, the upper set - the total masses of the
configurations. The maximal masses are marked with boxes.
}
\label{fig:M_density}
\end{figure}

As an illustration, I start with the study of ref.~\cite{Ippolito:2007hn}, 
where the problem of equilibrium and stability of non-rotating and rapidly 
rotating hybrid stars was addressed. This work describes the quark matter 
in the Nambu--Jona-Lasinio (NJL) model, which is a low-energy, 
non-perturbative approximation to QCD, that is anchored in the low-energy
phenomenology of the hadronic spectrum. While  dynamical symmetry
breaking, by which quarks acquire mass, is incorporated in this model,
it lacks confinement. A procedure which overcomes this difficulty in 
an {\it  ad hoc} manner is matching the quark equation of state to 
some low-density equation of state of hadronic matter. This can be 
done either via Gibbs or Maxwell construction; both constructions lead 
to quantitatively similar result, therefore we have adopted the latter.
In that case, at the deconfinement phase transition there is a jump 
in the density at constant pressure 
as illustrated in Fig.~\ref{fig:M_density},
left panel. The low-density nuclear equations of state are based on the 
Dirac-Bruckner-Hartree-Fock approach~\cite{WEBER_BOOK,Sedrakian:2006mq}.
The choice of a particular nuclear equation of state is 
dictated by the need to
match it to the quark equation of state. It turns out that it 
should be rather hard in order to fulfill this requirement
 Alternatively, one may consider the density at which the deconfinement 
takes place as a free parameter by adding an effective bag constant to 
the quark matter equation of state. In these, more phenomenological 
approaches, softer nuclear equations of state can be 
accommodated~\cite{Alford:2002rj}. 
If vector interactions are included in the NJL model
Lagrangian the stability of the models is improved and 
softer nuclear equations of state can be accommodated 
as well~\cite{Agrawal:2010er}.

The spherically symmetric solutions of Einstein's
equations for self-gravitating fluids are given by the well-known
Tolman-Oppenheimer-Volkoff equations~\cite{WEBER_BOOK,Sedrakian:2006mq}. 
A generic 
feature of these solutions is the existence of a maximum mass for 
any equation of state; as the central density is increased beyond 
the value corresponding to the maximum mass, the stars become 
unstable towards collapse to a black hole. A criterion for the 
stability of a sequence of configurations is the requirement
that the derivative $dM/d\rho_c$ should be positive (the mass should
be an increasing function of the central density). 
At the point of instability the fundamental (pulsation) modes 
become unstable. If stability is regained at higher central densities, 
the modes by which the stars become unstable towards the eventual 
collapse belong to higher-order harmonics.

For configurations constructed from a purely nuclear equation 
of state the stable sequence extends up to a maximum mass 
of the order 2 $M_{\odot}$ (Fig. \ref{fig:M_density}, right panel), 
this large value being a consequence of hardness of the equation of state.
The hybrid configurations branch off from the nuclear
configurations when the central density reaches that of the
deconfinement phase transition. It is seen that a stable branch 
of hybrid stars emerges in the range of central densities 
$1.3 \le \rho_c\le 2.5 \times 10^{15}$ g cm$^{-3}$. The masses of 
the quark cores cover the range  $0\le M_{\rm core}/M_{\odot}\le 0.75-0.88$ 
for central densities $1.3 \times 10^{15}\le \rho_c\le 2\times 10^{15}$. 
Thus, the quark core mass ranges from one third to about the half of the 
total stellar mass. 

\subsection{Rapidly rotating stars}

Millisecond neutron stars can rotate at frequencies which 
are close to the limiting orbital Keplerian frequency at 
which  mass shedding from the equatorial plane starts. 
The Keplerian frequency sets an upper limit on the rotation 
frequency, since other (less certain) mechanisms, such
as secular instabilities, could impose lower limits on
the rotation frequency. The mass versus central density 
dependence of compact stars rotating at the Keplerian frequency 
is similar to that for non-rotating stars  with the scales 
for mass shifted to larger values~\cite{Ippolito:2007hn}.
The increase in the maximum mass for stable hybrid configurations 
(in solar mass units) is $2.052\to 2.462$ for the model A,  
$2.017\to 2.428$ and $2.4174$ for the model A1 (there are two 
maxima) and  $1.981\to 2.35$ for the model B~\cite{Ippolito:2007hn}.

\begin{figure}[tb]
\includegraphics[height=.3\textheight,width=7.cm]{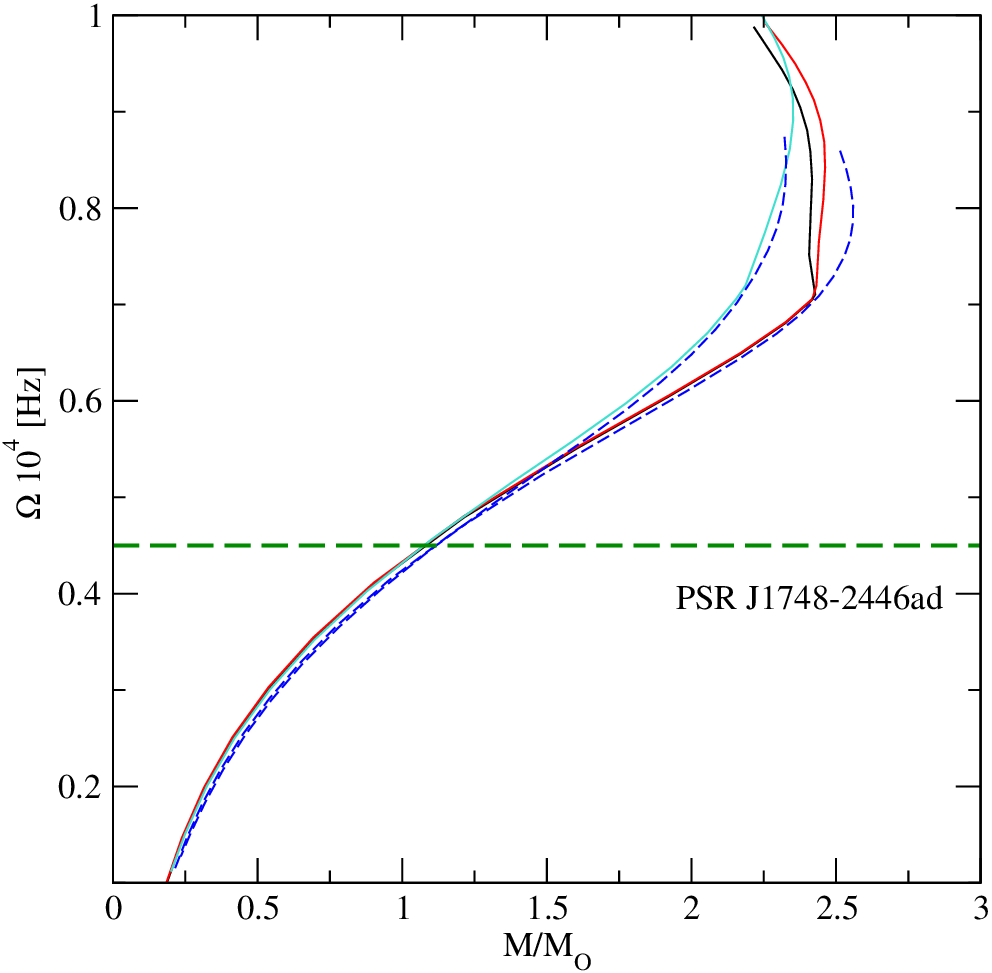}
\hskip 0.7cm
\includegraphics[height=.3\textheight,width=7.cm]{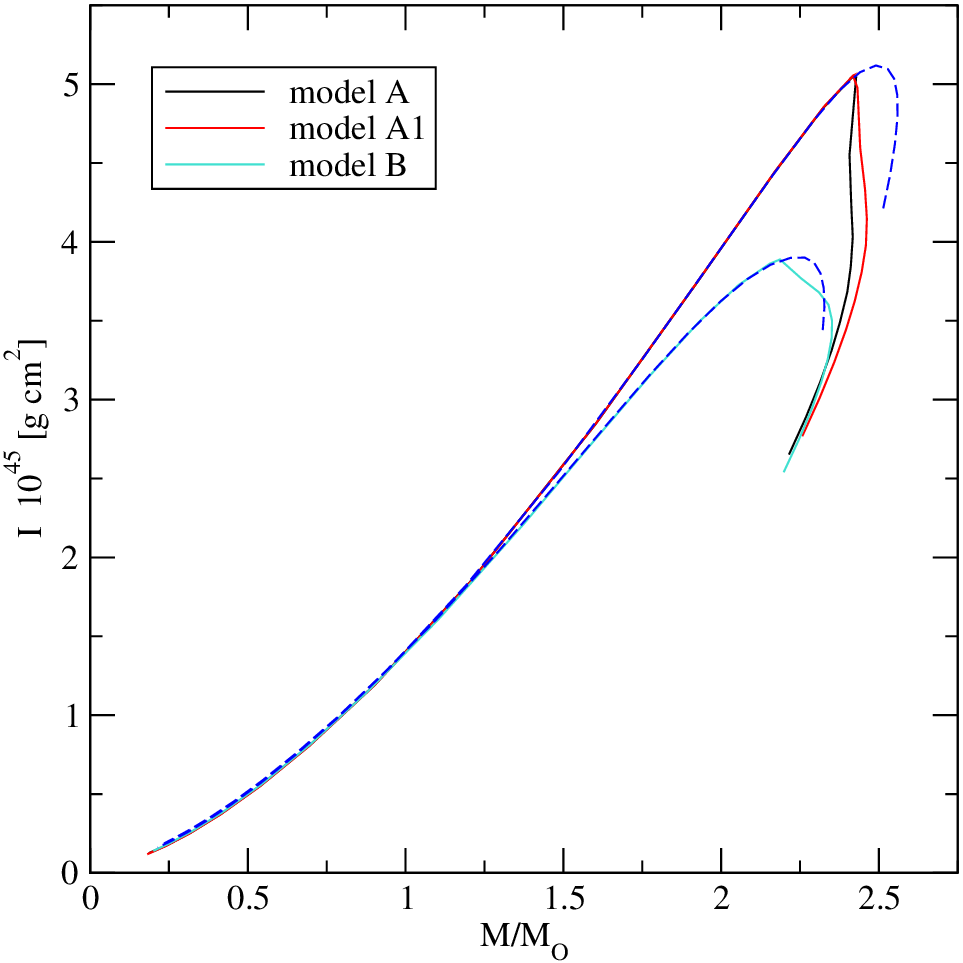}
\caption{
{\it Left panel}.
Dependence of Keplerian (maximal) frequency of rotation for
hybrid stars on their mass (in solar mass units). 
{\it Right panel}.
Dependence of moment of inertia of the hybrid configurations
on the mass of configuration (in solar mass units)
rotating at the limiting frequency. The conventions are the same
as in Fig.~\ref{fig:M_density}.
}
\label{fig:mass_om_kep}
\end{figure}

The recent discovery of the double-pulsar system 
PSR J0737-3039~\cite{Lyne,Burgay:2003jj}
offers a unique opportunity to place further bounds on the gross parameters
of compact stars by a measurement of the moment of inertia of star A, since
the spin frequencies and the masses of both pulsars are accurately measured.
Timing measurements over a period of years could provide  information
on spin-orbit coupling which could be revealed through an extra advancement
of the periastron of the orbit above the standard post-Newtonian advance
or in the precession of the orbital plane about the direction of the total
angular momentum~\cite{Lattimer:2004nj}.
The dependence of the moment of inertia, $I$, of configurations
on their mass (in the case of rotation at the Keplerian frequency) is shown
in Fig.~\ref{fig:mass_om_kep}. Since the moment of inertia is independent of the
rotation frequency, we have chosen to extract it for configurations rotating
at the limiting frequency; note that the masses of configurations should be
rescaled appropriately, if one is interested in the $I(M)$ function for slowly
rotating configurations. It is seen that 
while an accurate measurement of the moment of inertia of pulsar A can 
discriminate between the two nuclear equations of state, it will not be 
useful for accessing the properties of hybrid stars since the measured 
masses of pulsars are too low: $M/M_{\odot} = 1.337$ for pulsar A and 
$M/M_{\odot} = 1.250$ for pulsar B; as apparent from Fig.~\ref{fig:mass_om_kep} 
observations of heavier objects are needed to obtain useful bounds on 
the moment of inertia of hybrid configurations. The differences 
$\sim 20\%$ in the moments of inertia of purely nuclear and hybrid 
configurations of the same mass (in the case of the models A and A1) 
are within the accuracy that can be achieved in measurements similar to 
those carried out for PSR J0737-3039.

\subsection{Gravitational radiation}

It is interesting  that direct observational imprint of the size and 
mass distribution of the quark core will be carried by continuous 
gravitational waves if the core is 
crystalline-solid~\cite{Xu:2003xe,Owen:2005fn,Owen:2009tj} or 
superconducting-solid~\cite{Mannarelli:2008zz,Lin:2007,Haskell:2007,Knippel:2009st}.
Indeed, gravitational waves emitted by non-axisymmetric rotating compact 
stars are expected to be in the bandwidth of current gravitational wave 
interferometric detectors. Gravity waves arise from time-dependent 
quadrupole deformations of masses. In hybrid stars the axial symmetry 
of the star's core can be broken by the solid deformations in CCS matter.
Several authors estimated the strength of gravitational wave emission 
from crystalline-color-superconducting (CCS) 
matter~\cite{Lin:2007,Haskell:2007,Knippel:2009st}.
The characteristic strain amplitude of gravitational waves 
emitted by a triaxial star rotating about its principal axis is  
$
h_0=16\pi^2 G c^{-4} \epsilon I_{zz}\nu^2r^{-1},
$
where $\nu$ is the star's rotation frequency, $r$ is the distance to the 
observer, $\epsilon=(I_{xx}-I_{yy})/I_{zz}$ is the equatorial ellipticity, 
$I_{ij}$ is the tensor of moment of inertia, $G$ is the gravitational 
constant, $c$ is the speed of light. The elastic deformations are assumed 
to be small perturbation on the background equilibrium of the star. The 
current upper limit for the Crab pulsar $h_0 < 3.4 \times 10^{-25}$, 
which is rotating at the frequency $\nu = 29.6$ Hz  at the distance 
2 kpc~\cite{LIGO_S5_CRAB}. The key unknown parameter of the theory 
is the breaking strain of CCS core  $\bar\sigma_{\rm max}
\sim 10^{-3}-10^{-5}$ and the shear modulus, which can be computed 
approximately and scales as $\mu_{\rm shear}\sim \Delta^2$, 
where $\Delta$ 
is the pairing gap in CCS matter~\cite{Mannarelli:2008zz}. 
It follows then that $h_0 \propto \bar\sigma_{\rm max}\Delta^2$.
The strain of gravitational wave emission turns out to be 
close to the upper limit quoted above if $\Delta = 50$ MeV and 
 $\bar\sigma_{\rm max} = 10^{-4}$.  
The  Crab pulsar's current limit 
implies $\bar\sigma_{\rm max}\Delta^2  
\sim 0.25$ MeV$^2$, assuming maximally strained matter. 
The evolutionary avenues that may lead to such maximal 
deformations are not known. Pulsars may simply preserve 
their initial deformations as they cool down and solidify.
Additional contribution due to the deformations in the crust
of the star have not been taken into 
account in our estimates. However, this contribution is 
likely to relevant to low-mass neutron stars~\cite{Horowitz:2009vm}.

\section{Color-magnetic flux tubes }

The information gained from measurements of the 
gross parameters of compact objects, such as the
mass or the moment of inertia give us only indirect 
information on potential phases of matter in their 
interiors. The study of rotational, thermal and 
magnetic evolutions of neutron stars provide complementary
information, which in many cases is much more sensitive 
to the composition of matter. 

This section discusses the response of dense quark matter 
to magnetic field following Ref.~\cite{Alford:2010qf}. Our 
discussion will be focused on
the formation and dynamics of flux tubes in the 
two-flavor color superconducting (2SC) phase. 
The intact gauge symmetry in the 2SC phase  
is a linear combination of
the ordinary electromagnetic and color symmetries; this phenomenon 
is called ``rotated electromagnetism''~\cite{Alford:1999pb}.
The associated gauge field, the ``$\Qt$ photon'',
is a combination of the original photon and one of the gluons.
It is massless and propagates freely in 2SC quark matter.
The orthogonal combination $\X$ is a broken gauge generator, and
the associated magnetic field has a finite penetration depth.

Depending on the ratio of the $\X$-flux penetration depth to the
coherence length of the condensate, the 2SC phase may be
type-I or type-II with respect to the $\X$ magnetic 
field~\cite{Iida:2002ev,Sedrakian:2002ff}.
A superconductor is of type II if it obeys the condition
$
\kappa \equiv {\lambda}/{\xi} > {1}/{\sqrt{2}},
$
where $\kappa$ is the Ginzburg-Landau (GL) parameter,
$\lambda$ is the penetration depth, 
and $\xi$ is the coherence length of the superconductor.
For typical values of 2SC superconductor 
$
\kappa_{\rm 2SC} \approx 11{\Delta}/{\mu_q} \ .
$
Therefore, 2SC quark matter will be of type II if the 
pairing gap is sufficiently large, $\Delta \ge\mu_q/16$.
In quark matter we expect $\mu_q\sim 400$~MeV, 
so this only requires the 2SC pairing gap to be greater 
than about 25 MeV. Similar conclusions were 
reached in refs.~\cite{Iida:2002ev,Sedrakian:2002ff}.

Type-II superconductivity leads to formation of flux tubes 
if the cost of inserting a flux tube, which has always a positive 
self-energy (flux-tension) in a superconductor is compensated
by the interaction energy of the flux tube with external magnetic 
field intensity. The critical intensity (lower critical field) 
$H_{c1}$ for the formation of Abrikosov flux tubes containing 
$X$-magnetic flux is large, of order $10^{17}$ G.
However, we argued that, when the quark matter 
cools into the 2SC phase, the process of domain formation and
amalgamation is likely to leave some of the $X$ flux trapped 
in the form
of flux tubes. The exact configuration and density of such tubes
depends on details of the dynamics of the phase transition,
but the density could be within an order of magnitude of the
density of conventional flux tubes in proton-superconducting
nuclear matter. Note that the 
calculations apply to 2SC quark matter in the temperature 
range
$T_{1SC}<T \ll T_{2SC}$ where $T_{2SC}$ is the 
critical temperature for the formation of the 2SC 
condensate while $T_{1SC}$ is the critical temperature of 
formation of spin one condensate.

We shall assume that the flux tubes in the 2SC phase are topologically
stable. This is not obvious, since 
they are  analogous to ``$Z$-strings'' of the 
standard model~\cite{Vachaspati:1992fi} which have been found 
to be stable only in a small region of the standard model 
parameter space~\cite{James:1992wb}, although the stable 
region may be enlarged when
bound states are taken in to account~\cite{Vachaspati:1992mk}.
Because there are also differences between the
2SC phase of QCD and the Higgs phase of the standard model,
a separate stability calculation will be needed for the 2SC case.

The 2SC phase contains three species of gapless fermions: 
two quarks (``blue up'' and ``blue down'') and the electron. 
These are expected to dominate its transport properties.
Strange quarks can be included in the analysis 
as long as their pairing pattern does not
break the $\Qt$ gauge symmetry. 
Muons may also be present, but, like
strange quarks, their higher mass gives 
them a lower Fermi momentum
so they make a sub-leading contribution to
the transport phenomena.

Light ungapped fermions will be scattered from flux-tubes due 
to the Aharonov-Bohm effect; the cross-section of this process 
is given by~\cite{Alford:1988sj}
\begin{equation}
 \frac{d\sigma}{d\vartheta} = 
\frac{\sin^2(\pi\tilde\beta)}{ 2\pi k\sin^2(\vartheta/2)},
\label{AB-scattering}
\end{equation}
where
$
\tilde\beta= {q_p}/{q_c} \ ,
$
where $q_p$ is the charge of the scattering particle. 
For a flux tube that arises as a topological soliton 
in an Abelian Higgs model, $q_c$
is the charge of the condensate field whose winding 
by a phase of $2\pi$ characterizes the flux tube; 
$k$ is the momentum in the plane perpendicular to the 
string, and $\vartheta$ is the scattering angle.
Aharonov-Bohm scattering has several important features:
(i) The cross-section vanishes if $\tilde\beta$ 
is an integer, but is otherwise non-zero.
(ii) The cross section is {\em independent of the thickness
of the flux tube}: the scattering is not suppressed in the limit
where the symmetry breaking energy scale goes to infinity, and
the flux tube thickness goes to zero.
(iii) The cross section diverges both at 
low energy and for forward scattering.

The values of $\tilde\beta$ for scattering of the
fermions that are ungapped in the 2SC phase
off a flux tube containing magnetic flux
is given, to lowest order in the electromagnetic 
coupling constant $\alpha$, by
\begin{eqnarray}
\tilde\beta^\psi = {\rm diag}\Bigl(
  \dsp \frac{1}{2}+\frac{\alpha}{2\alpha_s},  
  \dsp \frac{1}{2}-\frac{\alpha}{2\alpha_s}, 
  \dsp \frac{1}{2}-\frac{\alpha}{2\alpha_s}, 
  \dsp \frac{1}{2}+\frac{\alpha}{2\alpha_s}, 
  \dsp -1+\frac{\alpha}{\alpha_s}, 
  \dsp -1,  
  \dsp -\frac{\alpha}{\alpha_s}  \Bigr) \ 
\label{ABfactor-approx}
\end{eqnarray}
in the basis defined as $\psi = (ru,gd,rd,gu,bu,bd,e^-)$,
where ``$ru$'' means the red up quark, etc. 
``$e^-$'' is the electron and $\alpha_s\sim 1$ is the strong 
coupling constant.
We conclude that the gapped quarks have 
$\tilde\beta$ close to $0.5$, which means
that they have near-maximal Aharonov-Bohm interactions with an $X$-flux tube.
Among the lighter (and hence more phenomenologically relevant) fermions,
the $\Qt$-neutral $bd$ has zero Aharonov-Bohm interaction with the flux tube,
while the $bu$ and electron have the same Aharonov-Bohm factor
$
\sin(\pi\tilde\beta^{bu})=\sin(\pi\tilde\beta^{e})
\approx -\pi (\alpha/\alpha_s) \ .
\label{ABfactor-bu}
$
The scattering of electrons and blue quarks of the flux tubes 
in the 2SC cores of compact stars contributes to the transport 
coefficients of matter  and acts as a dissipative force on the 
flux tubes. The relaxation time for incident light fermion
on a flux tube is given by~\cite{Alford:2010qf} 
\begin{equation}
\tau^{-1}_{if} = \frac{n_v}{p_{Fi}}  \sin^2(\pi\tilde\beta_i) \ .
\label{tauinv-flux}
\end{equation}
where $n_v$ is the flux density and $p_{Fi}$ is 
the fermion momentum. It is easy to understand the 
this result. It is of the standard form for classical 
gases $\tau^{-1}=c n \sigma$,  where $c=1$ is the speed of
the particles, $n=n_v$ is the density of scattering 
centers, and $\sigma\propto \sin^2(\pi\tilde\beta)/p_F$ 
is the cross section for Aharonov-Bohm scattering. 
One of the blue quarks, the blue down quark, has no 
A-B interaction with the flux tubes
($\tilde\beta=0$). The other two, the electron and 
blue up quark, have identical A-B factors 
\ref{ABfactor-bu} although their Fermi momenta are 
different.

Because the ambient magnetic field in a neutron star is below
the lower critical field required to force $X$-flux tubes into
2SC quark matter, the trapped flux tubes will feel a boundary 
force pulling them outwards.  The outward force per unit 
length on the flux tube is 
$
f_b =  r\epsilon_X/(R^2-r^2)
\label{fb-full}
$
where $\epsilon_X$ is the energy per length of $X$ flux tubes
(flux tension), $R$ is the radius of the 2SC core, $r$ is the 
radial coordinate in cylindrical coordinates. Other 
contributions to the magnetic field from outside of the 
2SC core have been neglected. Substituting 
the flux tension we obtain~\cite{Alford:2010qf}
\begin{eqnarray}
f_b \approx \frac{1}{3\pi}
\left(\frac{r}{R^2-r^2}\right)
\mu_q^2\ln\kappa_X \ ,
\label{fbmax}
\end{eqnarray}
where $\kappa_X$ is the Ginzburg-Landau parameter.
This force will be balanced
by the drag force (``mutual friction'') on the moving flux 
tube due to its Aharonov-Bohm interaction with the thermal 
population of gapless quarks and electrons, and also by 
the Magnus-Lorentz force. 
An estimate of the timescale for the expulsion of
$X$ flux tubes from a 2SC core on the basis of 
balance of forces acting on the vortex shows that 
it is of order $10^{10}$  years~\cite{Alford:2010qf}. 
Therefore, it is safe to assume that the magnetic field 
will be trapped in the 2SC core over evolutionary 
timescales in the absence of other external forces.

\section{Final remarks}

There has been substantial progress in understanding the 
astrophysics of compact objects containing quark matter 
in their interiors  over the past years. However much remains 
to be done. Further work is needed in developing methods to 
treat quark matter in non-perturbative regime relevant 
to compact stars. Finding its ground state in the cold 
and dense regime and within realistic models is another 
long-standing challenge. 
Astrophysics offers several avenues for discerning the state 
of the matter hidden inside compact stars. Some of them such
as measurements of their gross parameters, gravitational 
radiation and magnetism have been discussed above. Other 
sensitive tools provided by nature are the thermal cooling 
and anomalies in rotation of compact star. 
These issues are discussed elsewhere~\cite{Sedrakian:2009uu}.

\begin{theacknowledgments}

I thank M. Alford, N. Ippolito, B. Knippel, M. Ruggieri, 
D. H. Rischke, and F. Weber  for their contributions to the 
research described in this article and M. Alford for comments 
on the manuscript. I am grateful to the 
organizers of QCD@Work Giuseppe Nardulli memorial workshop
for their impressive efforts in arranging a 
successful conference. This work was supported, in part, 
by the Deutsche Forschungsgemeinshaft and by the European 
research netwrok ``Compstar''.

\end{theacknowledgments}

\bibliographystyle{aipproc}   
\bibliography{refs}

\begin{thebibliography}{99}
\bibitem{pairing_book}
 {\it Pairing in Fermionic Systems}, edited by A. Sedrakian, J. W. 
  Clark and M. Alford, (World Scientific, Singapore, 2006).



\bibitem{Bailin:1983bm}
  D.~Bailin and A.~Love,
  Phys.\ Rept.\  {\bf 107}, 325 (1984).

\bibitem{Alford:2007xm}
M.~G. Alford, A.~Schmitt, K.~Rajagopal, and T.~Schafer, 
Rev. Mod. Phys. {\bf 80}, 1455 (2008).


\bibitem{Casalbuoni:2003wh}
  R.~Casalbuoni and G.~Nardulli,
  Rev.\ Mod.\ Phys.\  {\bf 76}, 263 (2004).                           




\bibitem{Ippolito:2007hn}
  N.~Ippolito, M.~Ruggieri, D.~H.~Rischke, A.~Sedrakian and F.~Weber,
  Phys.\ Rev.\  D {\bf 77}, 023004 (2008).




\bibitem{WEBER_BOOK} F.~Weber,
{\it Pulsars as astrophysical laboratories for nuclear and particle 
physics},  Bristol, U.K.: Institute of Physics, 1999.

\bibitem{Sedrakian:2006mq}
  A.~Sedrakian,
  Prog.\ Part.\ Nucl.\ Phys.\  {\bf 58}, 168 (2007).

\bibitem{Alford:2002rj}
  M.~Alford and S.~Reddy,
  Phys.\ Rev.\  D {\bf 67}, 074024 (2003).

\bibitem{Agrawal:2010er}
  B.~K.~Agrawal,
  Phys.\ Rev.\  D {\bf 81}, 023009 (2010).

\bibitem{Lyne}
A. G. Lyne  et al., Science, {\bf 303},  1153 (2004).

\bibitem{Burgay:2003jj}
  M.~Burgay  et al.,
  Nature {\bf 426}, 531 (2003).

\bibitem{Lattimer:2004nj}
  J.~M.~Lattimer and B.~F.~Schutz,
  Astrophys.\ J.\  {\bf 629}, 979 (2005).


\bibitem{Xu:2003xe}
  R.~X.~Xu,
  Astrophys.\ J.\  {\bf 596}, L59 (2003).

\bibitem{Owen:2005fn}
  B.~J.~Owen,
  Phys.\ Rev.\ Lett.\  {\bf 95}, 211101 (2005).

\bibitem{Owen:2009tj}
  B.~J.~Owen,
  Class.\ Quant.\ Grav.\  {\bf 26}, 204014 (2009).


\bibitem{Mannarelli:2008zz}
  M.~Mannarelli, K.~Rajagopal and R.~Sharma,
  Prog.\ Theor.\ Phys.\ Suppl.\  {\bf 174}, 39 (2008).

\bibitem{Lin:2007}
L.-M.~Lin,  Phys.\ Rev.\  D {\bf 76}, 081502(R) (2007).


\bibitem{Haskell:2007}
B.~Haskell, N.~Andersson, D.~I.~Jones, and L.~Samuelsson,
Phys.\ Rev.\  Lett. {\bf 99}, 231101 (2007).


\bibitem{Knippel:2009st}
  B.~Knippel and A.~Sedrakian,
  Phys.\ Rev.\  D {\bf 79}, 083007 (2009).



\bibitem{LIGO_S5_CRAB}
B.~Abbott, {\it et al.} (LIGO Scientific Collaboration)
Astrophys. J. Letters {\bf 683}, 45 (2008).


\bibitem{Horowitz:2009vm}
  C.~J.~Horowitz,
  Phys.\ Rev.\  D {\bf 81}, 103001 (2010).



\bibitem{Alford:2010qf}
  M.~G.~Alford and A.~Sedrakian,
  J.\ Phys.\ G {\bf 37}, 075202 (2010).

\bibitem{Alford:1999pb}
 M.~G.~Alford, J.~Berges and K.~Rajagopal,
 Nucl.\ Phys.\  B {\bf 571}, 269 (2000).

\bibitem{Iida:2002ev}
 K.~Iida and G.~Baym,
 Phys.\ Rev.\  D {\bf 66}, 014015 (2002).

\bibitem{Sedrakian:2002ff}
  D.~M.~Sedrakian, D.~Blaschke, K.~M.~Shahabasian and D.~N.~Voskresensky,
  Phys.\ Part.\ Nucl.\  {\bf 33}, S100 (2002);
  Astrofiz.\  {\bf 44}, 443 (2001)
  [Astrophysics {\bf 44}, 359 (2001)].

\bibitem{Vachaspati:1992fi}
T.~Vachaspati, 
Phys.  Rev. Lett. {\bf 68,} 1977 (1992).

\bibitem{James:1992wb}
M.~James, L.~Perivolaropoulos, and T.~Vachaspati, 
Nucl. Phys. {\bf B395}, 534 (1993).

\bibitem{Vachaspati:1992mk}
T.~Vachaspati and R.~Watkins, 
Phys. Lett. {\bf B318}, 163 (1993).


\bibitem{Alford:1988sj}
  M.~G.~Alford and F.~Wilczek,
  Phys.\ Rev.\ Lett.\  {\bf 62}, 1071 (1989).

\bibitem{Sedrakian:2009uu}
  A.~Sedrakian,
  Acta Phys.\ Polon.\  B {\bf 3}, 669 (2010).


\end{thebibliography}

\end{document}